# Intersubband transition engineering in the conduction band of asymmetric coupled Ge/SiGe quantum wells

**Luca Persichetti[1], Michele Montanari[1], Chiara Ciano[1], Luciana Di Gaspare[1], Michele Ortolani[2], Leonetta Baldassarre[2], Marvin Zoellner[3], Samik Mukherjee[4], Oussama Moutanabbir[4], Giovanni Capellini[1,3], Michele Virgilio[5], and Monica De Seta[1,\*]**

1. Department of Sciences, Università Roma Tre, Viale G. Marconi 446, I-00146 Rome, Italy; luca.persichetti@uniroma3.it (L.P.); michele.montanari@uniroma3.it (M.M.); chiara.ciano@uniroma3.it (C.C.) luciana.digaspare@uniroma3.it (L.D.G.)
2. Department of Physics, Sapienza University of Rome, Piazzale Aldo Moro, I-00185 Rome, Italy; michele.ortolani@roma1.infn.it (M.O.); leonetta.baldassarre@roma1.infn.it (L.B)
3. IHP-Leibniz-Institut für innovative Mikroelektronik, Im Technologiepark 25, D-15236 Frankfurt (Oder), Germany; capellini@ihp-microelectronics.com (G.C.); zoellner@ihp-microelectronics.com (M.Z.);
4. Department of Engineering Physics, École Polytechnique de Montréal, Canada; samik.mukherjee@polymtl.ca (S.M.); oussama.moutanabbir@polymtl.ca (O.M.)
5. Department of Physics "Enrico Fermi", Università di Pisa, Largo Pontecorvo 3, I-56127 Pisa, Italy; michele.virgilio@unipi.it (M.V.)

\* Correspondence: monica.deseta@uniroma3.it; Tel.: +39- 06-5733-3430 (M.D.S.)

**Abstract:** *n*-type Ge/SiGe asymmetric-coupled quantum wells represent the building block of a variety of nanoscale quantum devices, including recently proposed designs for a silicon-based THz quantum cascade laser. In this paper, we combine structural and spectroscopic experiments on 20-module superstructures, each featuring two Ge wells coupled through a Ge-rich tunnel barrier, as a function of the geometry parameters of the design and the P dopant concentration. Through the comparison of THz spectroscopic data with numerical calculations of intersubband optical absorption resonances, we demonstrated that it is possible to tune by design the energy and the spatial overlap of quantum confined subbands in the conduction band of the heterostructures. The high structural/interface quality of the samples and the control achieved on subband hybridization are the promising starting point towards a working electrically pumped light-emitting device.

## 1. Introduction

Semiconductor multi-quantum wells (MQWs) represent the perfect playroom for nanoscale scientists to devise novel technologies and device architectures using the fundamental laws of quantum mechanics [1,2]. In the last decades, the number of applications based on a MQW structure core has been constantly expanding, and now ranges from photovoltaics and solar energy harvesting [3-5] to photonics [6-8] and spin-based optoelectronics [9,10]. Such a great technological potential exploits the tunability of the MQW design for precisely controlling the wavelength of electromagnetic radiation emitted or absorbed in the transitions between quantized states of the heterostructure. Compared to interband light emitters, unipolar devices based on intersubband (ISB) transitions in the conduction or valence band of the QW, such as quantum cascade lasers (QCLs) [11] and quantum fountains [12], offer a straightforward route to access the THz frequency domain and to extend the range of viable light-emitting materials to non-direct bandgap semiconductors [13,14]. As a matter of fact, since, in these unipolar devices, the radiative decay rate does not depend on electron-hole recombination, efficient light emitters can be envisaged with group-IV materials, such as Ge, SiGe alloys [14-16] and GeSn [17,18], which can be monolithically grown on silicon wafers and, thus, integrated in the complementary metal-oxide-semiconductor (CMOS) platform. Electrically pumped QCLs and optically pumped quantum fountains realized in the SiGe material system also benefit from the absence of polar longitudinal optical phonons inducing the long-range polarization field (Fröhlich interaction) which strongly couples to charge carriers and limits their non-radiative lifetimes in III-V compounds [7]. Being only mediated by the short-range deformation potential, electron-phonon coupling in non-polar semiconductors is much weaker [19,20] and, as a consequence, SiGe ISB lasers are expected to have a wider temperature operational range, reaching room temperature for *n*-type Ge/SiGe QCLs [21]. Also, the emission bandwidth would widen, extending inside the reststrahlen band (5-10 THz) of forbidden light propagation caused, in III-V materials, by strong light absorption by optical phonons in this spectral range [7]. The most promising QCL design for successful lasing operation in the group-IV material platform exploits ISB transitions of electrons in the conduction-band *L* valleys of Ge-rich heterostructures and features, as a building block, asymmetric-coupled quantum wells (ACQWs) made of Ge layers of different thicknesses and coupled through $Si_{(1-x)}Ge_x$ barriers, typically with a Ge content $x=0.77 \div 0.85$ [22,23]. This ACQW system inherently presents some of the main challenges for a future QCL development in the SiGe system [21], such as (*i*) achieving high structural and interface quality in strain-compensated Ge-rich heterostructures despite their large lattice mismatch with the Si substrate; (*ii*) control of the concentration profile of *n*-type dopant P atoms during the epitaxial growth and quantification of the profile broadening due to surface segregation of donors; (*iii*) designing of the energy of ISB transitions and engineering of electron wavefunctions with the desired degree of delocalization. As a necessary milestone on the way towards a SiGe based QCL, we investigate in this paper *n*-doped $Ge/Si_{0.20}Ge_{0.80}$ ACQWs grown by chemical vapor deposition (CVD), combining structural analysis techniques with THz spectroscopy as a probe of interwell coupling and level broadening due to interface roughness and ionized impurity scatterings. By modulating the heterostructure design (i.e. tunnel barrier thickness and relative width of the two QWs), we explore a wide range of ISB energies (different alignment of quantum-confined levels) and spatial distribution of the wavefunctions corresponding to excited confined states, obtaining a consistent match of the experimental data to numerical calculations performed through a multivalley effective mass model based on a Schrödinger-Poisson solver. This systematic study allows us to accurately determine material parameters for the SiGe system, such as ISB line broadenings and *L*-point conduction-band offsets, which are mandatory to establish a reliable simulation platform for QCL optimization.

## 2. Materials and Methods

Strain-compensated MQW heterostructures are grown by ultra-high vacuum CVD at 500 °C using ultrapure germane and silane without carrier gases. The reacting gas pressure is 1.2 mTorr, at which, the typical growth rate is 6.5 nm/min for Ge and 4.5 nm/min for a $Si_{0.20}Ge_{0.80}$ alloy. The sample design (sketched in Fig. 3c) features a wide and a narrow Ge well of thickness $w_L$ and $w_t$, respectively. The wells are separated by a SiGe tunnel barrier of thickness $b_t$ and Ge composition $x_{Ge}$. Among different samples, $w_t$= 5.0 nm is kept constant, $w_L$ is varied between 11.3 and 16.0 nm and $b_t$ ranges between 2.3 and 5 nm. $x_{Ge}$ has been set to 0.81 or 0.87, with the higher Ge composition describing the more pronounced SiGe intermixing for $b_t<$ 3 nm [21]. The wide well is *n*-doped by phosphine codeposition over a thickness $t$= 10 nm. In the samples, the module, composed by the two Ge wells and the SiGe barrier separating them, has been repeated 20 times with 21 nm-thick $Si_{0.20}Ge_{0.80}$ spacers between the individual modules. Specific information on individual samples can be found in Table 1. The MQW stack is deposited on a relaxed $Si_{0.15}Ge_{0.85}$ alloy buffer with a thickness of 1.2 μm. In such high Ge content regime, low threading dislocation density (TDD~ $1\times10^7$ cm$^{-2}$) is obtained through a reverse step-graded virtual substrate (RG-VS) [24]. In this approach, first, a plastically relaxed 700 nm-thick Ge buffer is directly deposited on the Si(001) substrate and then, on top of it, two SiGe layers are deposited, each being 150 nm thick and with a Ge content, respectively of 0.95 and 0.90 (See schematics in Fig. 1a). Structural characterization of the samples is performed by scanning transmission electron microscopy (STEM) using a FEI Titan microscope, operated at 200 kV and equipped with aberration-corrected magnetic lenses for obtaining electron probes of the order of 1-2 Å diameter with a beam currents of 200 pA. CEOS CESCOR corrector was used to yield a resolution of 0.8 Å. The images were recorded using a high-angle annular dark field (HAADF) detector. The sample preparation for STEM was done in a dual-channel focused-ion-beam (Dual-FIB) microscope, using the standard lamella lift-out technique. STEM measurements are coupled to high-resolution X-ray diffraction (XRD) analysis performed at room temperature with a Rigaku SmartLab tool with a rotating anode and line-focus geometry featuring a Ge(400)x2 channel-cut beam collimator and a Ge(220)x2 analyzer crystal. The typical broadening of spatial donor profiles due to segregation and diffusion of P atoms in our growth conditions is measured on calibration samples using dynamic secondary ion mass spectrometry (D-SIMS) on a CAMECA IMS Wf Tool, with oxygen at 400 eV impact energy in positive mode, monitoring the 31P+ signal. ISB absorption spectra have been measured at $T$= 10 K by means of Fourier-Transform infrared (FTIR) spectroscopy in a side-illuminated single-pass waveguide configuration with a Bruker Vertex 70v equipped with a He-flow cryostat. The lateral facets of our 2.5 mm long samples are cut at a 70° angle with respect to the growth plane and the top surface close to the MQWs is coated with a metal bilayer (Ti/Au 10 nm/ 80 nm) [25], in order to make the electric field of the radiation propagating through the MQWs almost parallel to the ISB dipole moment (i.e. TM polarized) [26]. The measured quantity is the dichroic transmission spectra $T(\omega)=T_{TM}(\omega)/T_{TE}(\omega)$, which ensures that polarization independent spectral features not related to ISB transitions (e.g. the dopant absorption in the Si wafer) are suppressed. From $T(\omega)$, we evaluate the dimensionless absorption coefficient $\alpha_{2D}(\omega)$ and the sheet carrier density as in Ref. [27]. In the simulations, electron states and ISB absorption spectra have been calculated self-consistently in a Schrödinger-Poisson iterative scheme with parabolic subband dispersion [27], in which depolarization shift effects, i.e. the blueshift of absorption peaks due to the screening of the radiation field by the collective plasma mode [25], are included.

**Table 1.** ACQW design. $b_t$ is the thickness of the tunnel barrier and $x_{Ge}$ its composition. $w_L$ is the thickness of the wide well which is *n*-doped, except for sample 2223. For all the samples, the thin well is 5 nm-thick, the ACQW module is repeated 20 times and the $Si_{0.20}Ge_{0.80}$ spacer thickness between the modules is 21 nm. The MQW periodicity $D^{SL}$ is measured from the spacing of the superlattice fringes in XRD rocking curves. Values in brackets are the nominal periodicities. The sheet carrier density $n_{2D}$ is obtained from the FTIR optical absorption spectra $\alpha_{2D}(\omega)$.

| Sample | $w_L$ (nm) | $b_t$ (nm) | $x_{Ge}$ | $D^{SL}$ (nm) | $n_{2D}$ (x$10^{11}$ cm$^{-2}$) |
|---|---|---|---|---|---|
| 2216 | 12.0 | 2.3 | 0.87 | 41.6 [40.3] | 7.2 |
| 2217 | 13.0 | 2.3 | 0.87 | 42.8 [41.3] | 9.0 |
| 2218 | 11.3 | 2.3 | 0.87 | 40.6 [39.6] | 7.8 |
| 2219 | 11.5 | 2.3 | 0.87 | 40.6 [39.8] | 0.9 |
| 2221 | 12.0 | 3.3 | 0.81 | 42.4 [41.3] | 6.0 |
| 2222 | 12.0 | 4.0 | 0.81 | 43.0 [42.0] | 4.6 |
| 2223 | 12.0 | 2.3 | 0.87 | 39.5 [40.3] | undoped |
| 2224 | 12.0 | 5.0 | 0.81 | 43.9 [43.0] | 1.5 |
| 2267 | 16.0 | 2.3 | 0.81 | [44.3] | 3.0 |

## 3. Results

*3.1. Structural characterization*

Figure 1 shows a full STEM characterization of a typical ACQW sample, probed at different length scales and depths along the VS and the overgrown MQW stack. Taking as a reference the sample structure sketched in Fig. 1a, we fully characterize the grown stack, obtaining a high-resolution imaging of the active region composed by the 20 repeating ACQW modules (Fig. 1b-d), as well as probing the RG-VS and the Ge buffer layer (Fig. 1e) to characterize plastic relaxation at the bottom of the of 1.2 µm-thick $Si_{0.15}Ge_{0.85}$ alloy buffer. The images displayed in panels c and d show a zoom-up of the well/spacer (Ge/$Si_{0.20}Ge_{0.80}$) and the well/tunnel barrier (Ge/$Si_{0.13}Ge_{0.87}$) heterointerfaces where atomic-resolved features are evident. Between the thick $Si_{0.15}Ge_{0.85}$ alloy and the Ge buffer in Fig. 1e, we clearly observe three interfaces corresponding to the 5%-steps in the Ge concentration decreasing from pure Ge to $Si_{0.15}Ge_{0.85}$. The STEM contrast is due to the misfit dislocations that pile up at each interface. Threading dislocation segments are only present in the relaxed Ge buffer and in the lower RG layers, whereas no threading dislocation are observed in the top $Si_{0.15}Ge_{0.85}$ alloy buffer. The absence of threading segments within the field of view of the image is consistent with the TDD count reported in the methods and makes visually clear the "filtering" effect produced by the heterointerfaces which hinder the propagation upward of threading dislocations [25]. At the origin, there is the effective gliding force due to the compositional change at the interface which drives the change of the character of dislocations from threading to misfit, as threading segments bend on the interface plane [28,29]. Focusing on the quality of the MQW stack, large-scale STEM images (an example is reported in Fig. 1b) demonstrate the remarkably high reproducibility in the deposition process, resulting in a homogeneous periodicity along the growth direction. At high magnifications (Fig. 1c,d), STEM shows sharp and abrupt interfaces, with a broadening due to SiGe intermixing in the order of 0.8 nm [21]. The quality of the heterointerfaces is confirmed by the low value of the root-mean-square interface roughness which was estimated to be 0.18 nm by atomic probe tomography measurements performed on the same set of samples [30]. The thickness of wells and barriers reported in Table 1 was measured by energy-dispersive x-ray spectroscopy (EDX), while the periodicity of the MQW stack $D^{SL}$ was obtained by XRD. The latter provides the perfect tool for evaluating the structural quality and homogeneity at a length-scale of a few millimeters (i.e. the typical size of the x-ray beam), thus giving access to a spatial probe complementary to STEM.

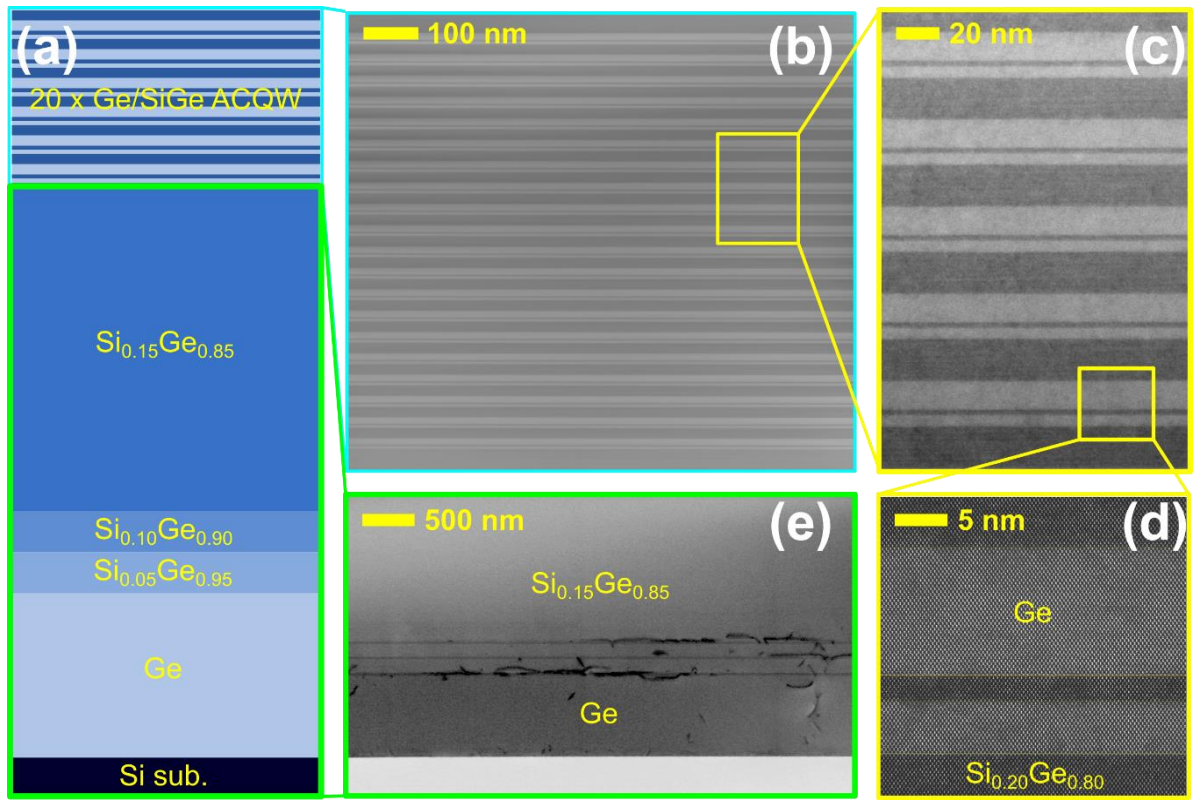

**Figure 1.** (a) Schematics of the structure of the grown heterolayers from the VS to the active region on top. (b-e) STEM micrographs of a typical ACQW sample: panels (b)-(d) show increasing magnifications of the active region composed by the alternation of Ge wells and SiGe barriers at different compositions, as described in the text. Panel (e) reports the part of the VS underlying the active region being close to the Si substrate, including the Ge buffer and the RG SiGe step layers.

Figure 2a shows a typical rocking curve, around the (004) Ge and (004) Si Bragg peaks, obtained on our ACQWs. As an exemplificative example, we report data for sample 2223. Together with the reflections due to the Ge buffer and the Si substrate, we observe the peak corresponding to the $Si_{0.15}Ge_{0.85}$ alloy buffer which is the thicker layer in the RG-VS. In addition, multiple orders of superlattice (SL) satellites emerge. Their high quality-factor confirms the high crystalline quality and sharpness of the MQW layers. A statistical analysis performed on the data reported in Table 1 for the entire set of samples reveals that the superlattice periodicity $D^{SL}$ matches the nominal values within 2.4% on average, with the maximum deviation, observed for the highest doping concentration ($n_{2D}$= 9x $10^{11}$ cm$^{-2}$), remaining below 3.6%. Information extracted from the rocking curve are paralleled by the reciprocal space map (RSM) around asymmetric (224) reflections shown in Fig. 2b for the same sample. From the map, we obtain a clear picture of the strain conditions in the Ge and SiGe layers. By taking as a reference the dashed diagonal line, which corresponds to fully relaxed SiGe alloys, it is visually clear that both the Ge and the $Si_{0.15}Ge_{0.85}$ spots do not lie on this line, being, therefore, tensile strained in the growth plane with $\varepsilon_{//}$= 0.16% and 0.19%, respectively. Therefore, despite being plastically relaxed, the RG-VS is not fully relaxed due to the contribution of thermal strain, unavoidably arising from the difference in the thermal expansion coefficients of Si and Ge, and which builds up upon cooling down the samples to room temperature after the growth [31]. In addition, we note the perfect vertical alignment of all the SL fringes with respect to the $Si_{0.15}Ge_{0.85}$ buffer, indicating that the entire MQW stack is coherent with the in-plane lattice parameter of the underlying VS.

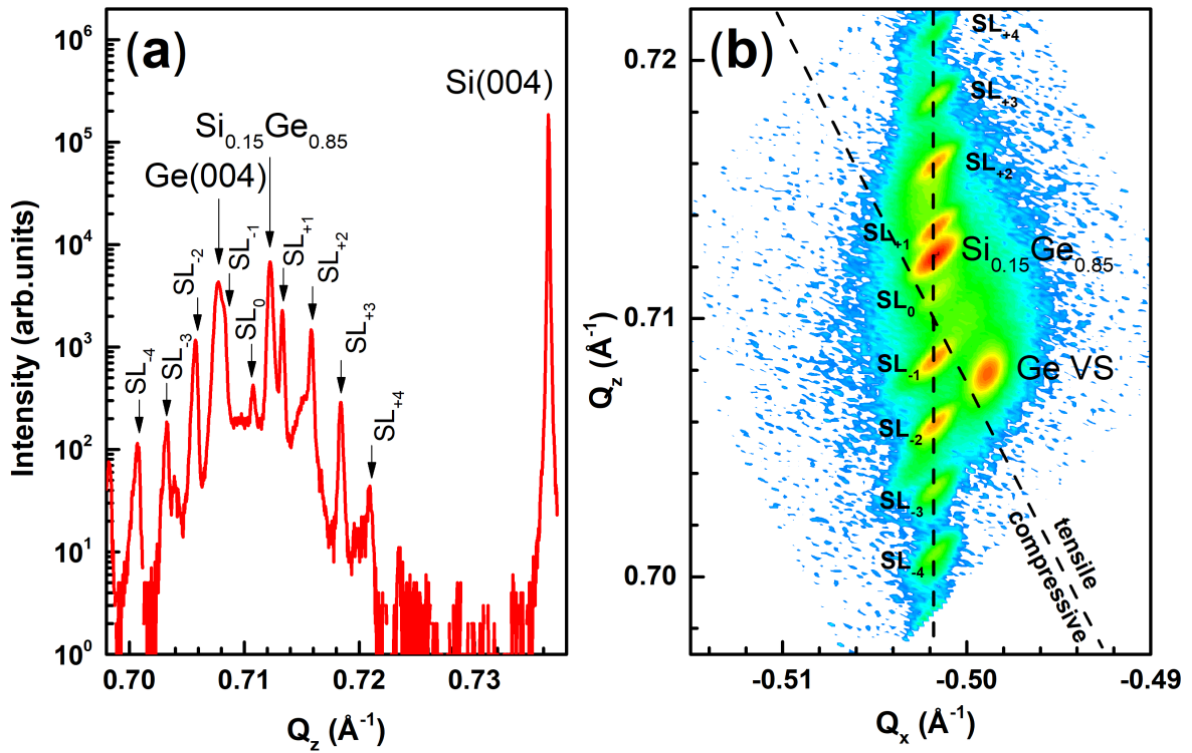

**Figure 2.** (a) XRD rocking curve around the (004) reflection and (b) XRD RSM around the (224) reflections of the SiGe sample 2223. We labelled the peaks associated to the Si substrate, the Ge buffer and the thick $Si_{0.15}Ge_{0.85}$ alloy buffer of the VS, as well as different orders of the SL fringes.

We now focus on the SIMS calibration of the $N_{3D}$ donor density in Ge films obtained by phosphine co-doping and its correlation to the actual carrier density $n_{2D}$ measured by FTIR. To this end, we grew a set of Ge/SiGe MQWs (20 periods) where the $PH_3$ partial pressure $p_{PH3}$ was varied within a wide range of values. To be meaningful for $PH_3$-$GeH_4$ co-depositions, such values of pressure need to be normalized to the $GeH_4$ flux $\phi_{GeH4}$, defining a reference parameter $P_{PH3}= p_{PH3}/ \phi_{GeH4}$ in units of mTorr/sccm. This quantity is related to the measured $N_{3D}$ in Fig. 3a (red dots), showing that $n$-type doping of Ge is doable till the high $10^{19}$ $cm^{-3}$ range where we observe a plateau in the donor density. Note that SIMS detection of $N_{3D}< 2\times10^{17}$ $cm^{-3}$ are particularly challenging in the SiGe system due to crosstalk of the 31P+ signal with hydrogenated Si species and, thus, data are sparser in this range. On the samples with a doping range measurable by FTIR and relevant for optically active MQW stacks, we compared donor detection by SIMS to the estimation of $n_{2D}/t$ (blue squares) where the sheet carrier density is obtained from optical absorption measurements. We find a good match between the two datasets, with the doping density estimated by FTIR being only slightly lower than the SIMS values. This indicates that carriers resulting from donor ionization effectively populate the confined ground state at the $L$ point in the well. Or, in other words, the density of defects in our samples is low enough not to evidence, at the doping densities explored, a mismatch due to the compensation of the $p$-type background arising from electrically charged defects (i.e. dislocations).

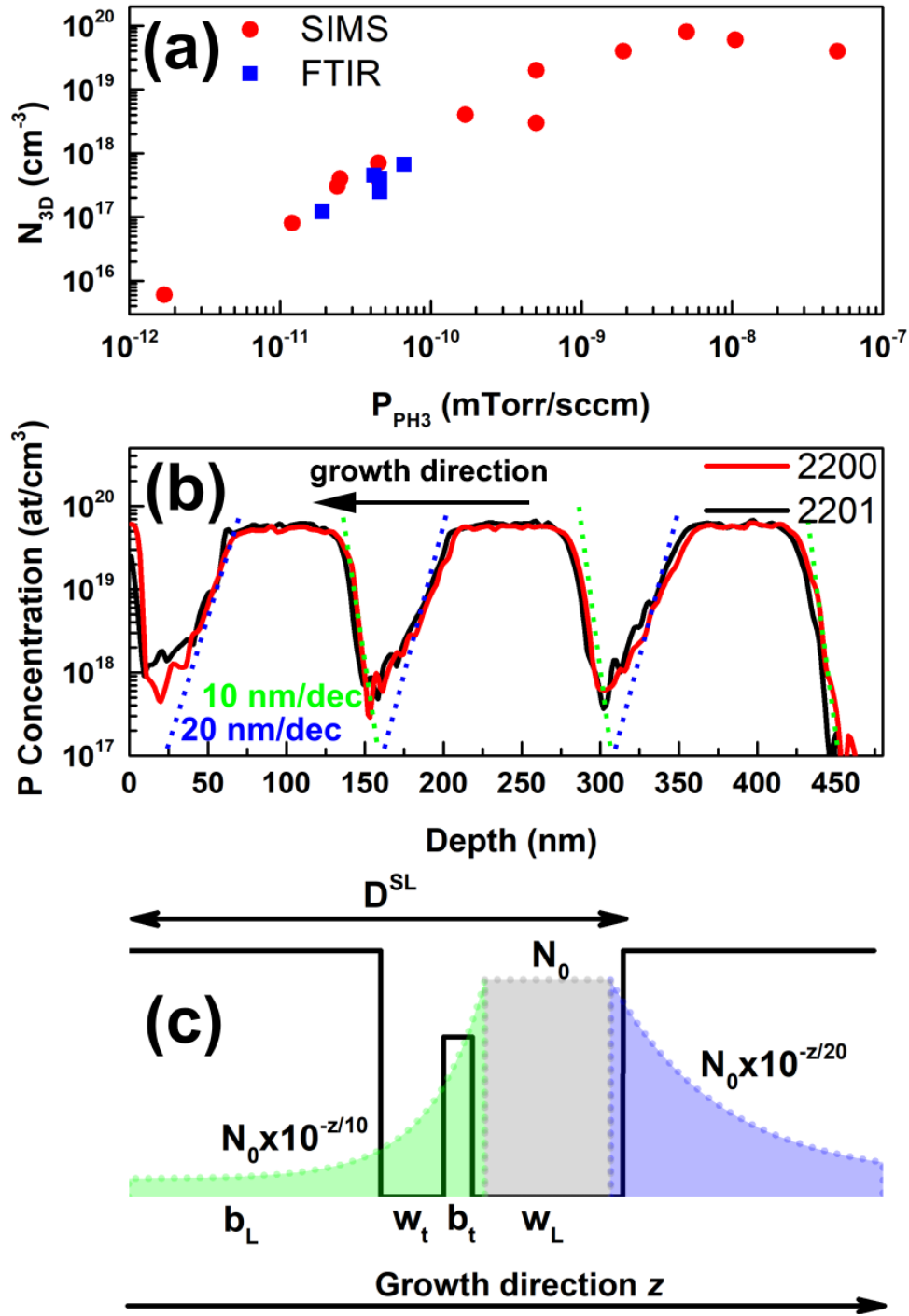

**Figure 3.** (a) (red dots) Correlation between the $N_{3D}$ concentration of P dopants obtained by SIMS and the growth parameter $P_{PH3}$, i.e. the phosphine partial pressure normalized to the flux of the codeposited germane gas. (blue squares) The same correlation plot but with the $N_{3D}= n_{2D}/t$ values obtained from the sheet carrier density $n_{2D}$ measured by FTIR and the thickness $t$ of the doped layer. (b) Spatial profile of P dopants along the growth direction z obtained by SIMS on two calibration samples featuring alternating doped and intrinsic layers. The diffusion tail of P atoms for the doped/undoped (undoped/doped) interfaces have been fitted to a mono-exponential decay function $N_0 \times 10^{-z/d}$ displayed as green (blue) dots, where $d$ is the characteristic decay length and $N_0$ the $N_{3D}$ density inside the doped layer. (c) Schematics of the ACQW module. The geometric parameters mentioned in the text are indicated. In the exponent of the decay function, $z$ is in nanometer. The donor diffusion in the leading and trailing directions is pictured.

A realistic modelling of ISB optical absorption demands to accurately input not only the absolute value of the doping density but also how the donors are spatially distributed along the growth direction. This is obtained through a spatially resolved mapping of the dopant P atoms by SIMS on calibration samples (Fig. 3b). Following the direction of the growth (black arrow in Fig. 3b), at the interfaces between the doped/undoped regions, we observed a leading edge-out diffusion of the donors with an exponentially decaying tail of 20 nm/decade (violet tail in Fig. 3c). Diffusion in the trailing edge of the reversed undoped/doped interface (green tail) is narrower (10 nm/decade), suggesting that the segregation of dopants occurring simultaneously as the growth proceeds is dominant over dopant diffusion at later stages of the growth. The obtained doping profile is schematically superimposed to the ACQW design in Fig. 3c, showing that, even in the growth configuration where the doped well is grown at the end of the repeated module, a non-negligible density of donors is found inside the tunnel barrier layer. This significantly affects the energy position of the subband minima and, therefore, the optical absorption spectra discussed in the following.

*3.2. Optical absorption measurements*

From the experimental absorption coefficient $\alpha_{2D}(\omega)$ (shown in Fig. 4a-d), we can extract both the linewidth and the energy position of ISB transitions for different design parameters. The corresponding theoretical spectra, calculated by the Schrödinger-Poisson solver, are reported in panels e-h. Focus first on samples 2218 and 2219 (Figs. 4a,b and 4e,f): The two samples share the same geometrical parameters, but their doping concentrations differ by almost one order of magnitude (see Table 1). As detailed in the following, their common design ($w_L \simeq$ 11.5 nm, $b_t$= 2.3 nm) targets a significant interwell coupling and, as a result, the absorption spectra of both the samples at *T*= 10 K clearly show two absorption resonances, centered at $E_{01}^{abs}$ and $E_{02}^{abs}$, with similar oscillator strengths. These resonances correspond to transitions from the ground- (*E₀*) to the first-excited (*E₁*) and the second-excited (*E₂*) subbands of the ACQW system, respectively. Reminding that in symmetric systems a single ISB absorption peak is visible at low temperature, the appearance of both the resonances in 2218 and 2219 demonstrates instead a significant overlapping of the envelope wavefunctions of the two wells. This is confirmed by the simulations, which match very well the experimental spectra in terms of energy and spectral weight of the two resonances. In particular, we note that the energy position of the absorption resonances in 2218 is significantly blue shifted with respect to those in 2219, due to the higher depolarization shift effect resulting from the heavier doping concentration of 2218 ($n_{2D}$= 7.8$x$10¹¹ cm⁻², instead of 0.9$x$10¹¹ cm⁻² measured on 2219). In addition, figure 4i shows that the doping concentration has a markedly pronounced impact on both the $\Gamma_1$ and $\Gamma_2$ absorption linewidths, evaluated, from a Lorentzian fit to the data, as the half-width at half-maximum (HWHM) values of the absorption peaks at $E_{01}^{abs}$ and $E_{02}^{abs}$, respectively. Data represented as hexagonal markers correspond to a set of heavily doped samples ($n_{2D}$> 7$x$10¹¹ cm⁻²) with increasing $w_L$ at constant tunnel barrier thickness $b_t$= 2.3 nm. For $w_L \simeq$ 11.5 nm, the lighter doped sample 2219 shows significantly smaller absorption linewidths, see square markers in Fig. 4i. We suggest that the observed narrowing of linewidths is driven by the lower impact of ionized-impurity scattering when the doping level is reduced from 7 to 0.9 $x$10¹¹ cm⁻² [32]. By further analyzing Fig. 4i, we note a strong reduction of $\Gamma_2$ when $w_L$ is increased. Its origin can be traced back to the peculiar shape of the envelope wavefunctions of the excited states and it will be discussed in the following section.

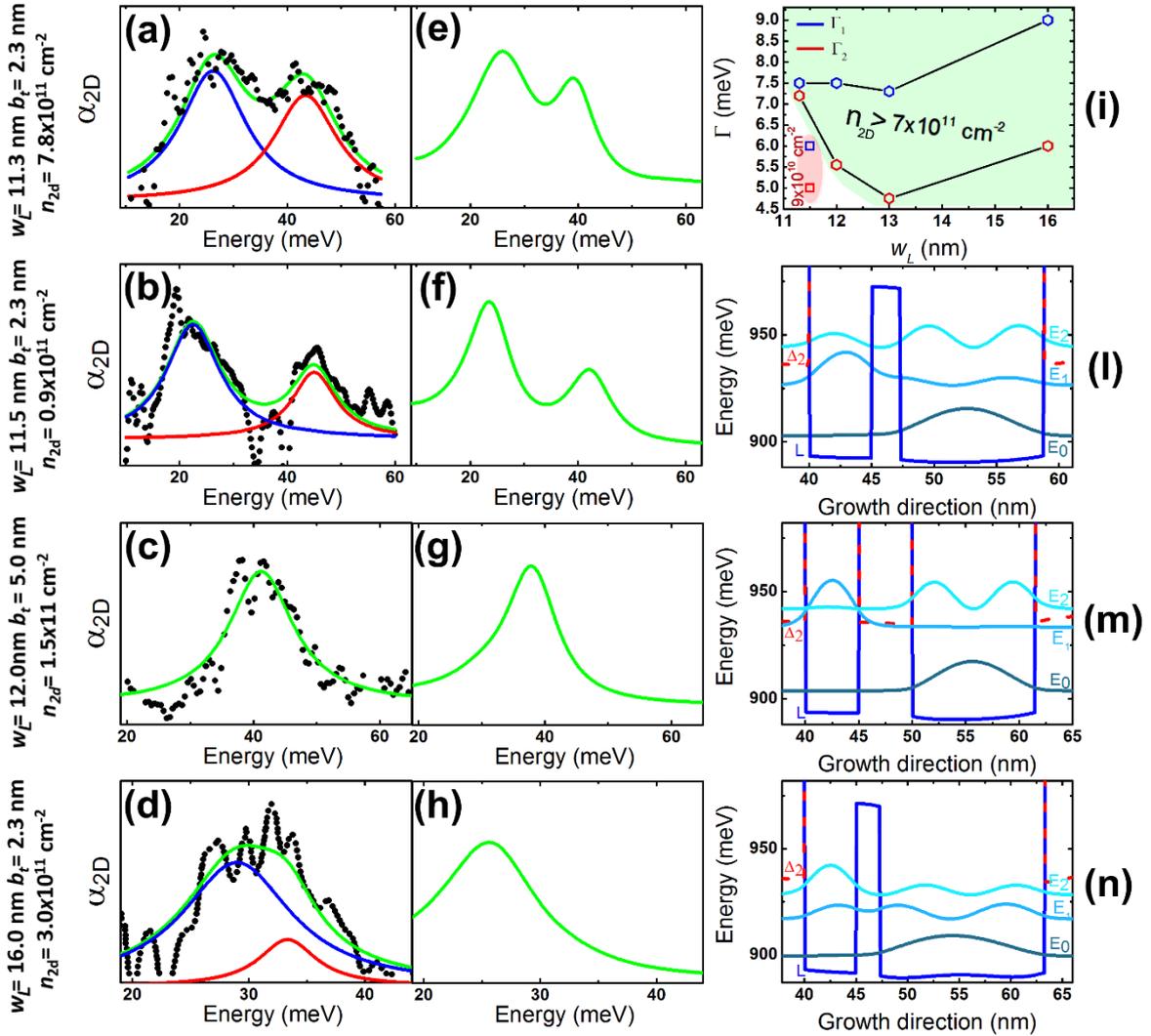

**Figure 4.** (a-d) Experimental ISB absorption measured by FTIR (black dots) with Lorentzian fits of the resonances centered at $E_{01}^{abs}$ (blue curve) and $E_{02}^{abs}$ (red curve). The green curve is the convolution of the blue and red fit curves. (e-h) Corresponding calculated absorption spectra. (a, e) Sample 2218; (b, f) Sample 2219; (c, g) Sample 2224; (d, h) Sample 2267. (i) Absorption linewidths $\Gamma_1$ and $\Gamma_2$ as a function of $w_L$. The values are the HWHM of the absorption peaks at $E_{01}^{abs}$ and $E_{02}^{abs}$, respectively, obtained as a Lorentzian fit to the experimental data. The green shaded area highlights the linewidths of the samples 2216, 2217, 2218, 2267 doped to $n_{2D}> 7x10^{11}$ cm$^{-2}$ (hexagonal markers). The red shaded area marks the linewidths obtained on sample 2219 (squares) for which $n_{2D}= 0.9x10^{11}$ cm$^{-2}$. (l-n) Calculated squared envelope wavefunctions for sample (l) 2219, (m) 2224, (n) 2267.

We now correlate the geometrical parameters of the ACQW design to the features observed in ISB optical absorption spectra. The coupling of the excited electronic states of the two wells, which has a direct impact on the spectral features, is tunable by changing either the thickness of the tunnel barrier or the relative width of the two wells. In Fig. 4, both the approaches are presented with exemplificative experimental absorption spectra compared to their numerical counterparts. The effect of increasing the thickness of the tunnel barrier is clear when comparing the results obtained on samples 2218 and 2224 and shown in Figs. 4(a) and 4(c), respectively. These two samples have a similar well width while $b_t$ changes from 2.3 nm (Sample 2218) to 5.0 nm (sample 2224). Hence, the $E_1$ envelope wavefunction in 2224 is strongly localized inside the thin well with a negligible penetration into the wide well, as shown by the calculated squared envelope wavefunctions in Fig. 4m. The spatial overlapping of $E_0$ and $E_1$ is therefore vanishing and a single absorption resonance is observed in both the experimental and calculated spectra of 2224, since the 0→1 oscillator strength is negligible. Sample 2267 provides, instead, an exemplificative case of wavefunction engineering by tuning $w_L$. The related absorption experiments and simulations are shown in Figs. 4d,h. With respect to samples 2218 and 2219 where $w_L \simeq 11.5$ nm, here, the wide well width is increased to 16.0 nm, while the tunnel barrier thickness is kept constant at $b_t = 2.3$ nm. By comparing the calculated squared wavefunctions of 2267 (Fig. 4n) to those obtained for 2219 in Fig. 4l, we note that their shapes for the excited states differ significantly, in particular for $E_2$. As demonstrated by the simulation of the optical absorption coefficient $\alpha_{2D}(\omega)$ of 2267 displayed in Fig. 4h, the modified symmetry of the envelope wavefunctions drives a sizeable change in the spectral weights of the two ISB resonances, with the contribution centered at $E_{02}^{abs}$ being markedly lower than in 2218 and 2219. More insights on such strong link between geometry and electronic/optical properties of ACQW systems can be gained by analyzing in detail the results of the numerical model in the next section.

## 4. Discussion

The blue (red) solid curve in Fig. 5 shows the calculated ISB transition energy $E_{01} = E_1 - E_0$ ($E_{02} = E_2 - E_0$) between the fundamental and the first- (second-) excited electron state of the ACQW system as a function of $w_L$ at a fixed $b_t = 2.3$ nm and $n_{2D} = 7.0 \times 10^{11}$ cm$^{-2}$. The corresponding values predicted for the optical resonance energy $E_{01}^{abs}$ ($E_{02}^{abs}$) are shown as a dotted blue (red) curve. The energy mismatch between ISB transitions and corresponding optical resonances is due to the depolarization shift effect, which is fully accounted in the calculated spectra shown in Fig. 4. The experimental optical absorption energies measured by FTIR on the samples featuring $b_t = 2.3$ nm and $n_{2D} > 7 \times 10^{11}$ cm$^{-2}$ are displayed by empty hexagonal markers, while the squares represent the optical absorption energies of the lower doped sample 2219. The experimental datapoints of the higher doped samples show a clear monotonic dependence with $w_L$, reproducing the behavior predicted by the model. As expected, the lower doped sample, deviates from the hexagonal dataset, due to its lower depolarization shift; the deviation is appreciable, in particular, for $E_{01}^{abs}$ on which the depolarization correction is higher [25].

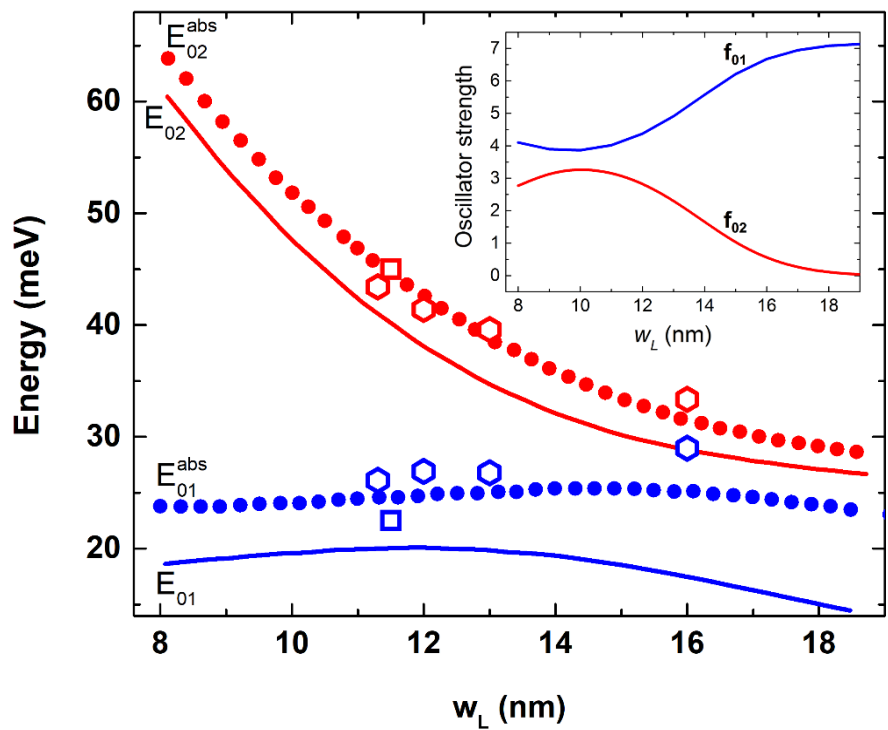

**Figure 5.** (a-d) Calculated energies for ISB transitions (continuous lines) and optical resonances (full dots) for ACQW structures with a narrow well of 5 nm and a Si$_{0.13}$Ge$_{0.87}$ tunnel barrier $b_t = 2.3$ nm as a function of the wide well width $w_L$. The blue (red) color code represents the ISB transition energy

$E_{01}$ ($E_{02}$) and the related absorption resonance energy $E_{01}^{abs}$ ($E_{02}^{abs}$). In the simulations, the sheet carrier density is $n_{2D}$= 7x10$^{11}$ cm$^{-2}$. The hexagonal (square) empty markers represent the experimental resonance energies obtained from FTIR spectra for samples with $n_{2D}$> 7$x$ 10$^{11}$ cm$^{-2}$ ($n_{2D}$= 0.9$x$ 10$^{11}$ cm$^{-2}$). In the inset, the calculated oscillator strength $f_{01}$ ($f_{02}$) for the 0->1 (0->2) transition is displayed as a function of $w_L$ for the same design parameters as in the main panel.

From the results of the calculations, we also note that the minimum separation between $E_{01}$ and $E_{02}$ is predicted to be around $w_L$ =14 nm which is the anticrossing point where $E_1$= $E_2$, in the limit of non-interacting wells (i.e. large tunneling barrier). The opening of a gap at the anticrossing reflects the hybridization between the states which, in the limit of uncoupled wells, represent the ground state in the narrow well and the first-excited state in the wide well. From the inset of Fig. 5, where we show the oscillator strength calculated for the 0->1 (0->2) ISB transition as a function of $w_L$, the degree of state hybridization is evidently higher for $w_L$< 12 nm. In this range, in fact, the two ISB transitions feature comparable oscillator strengths. By looking back at the spectroscopy data of Fig. 4, this explains the similar spectral weights of the optical resonances for designs with $b_t$= 2.3 nm and $w_L \simeq$ 11.5 nm, see panels (a and b). The significant state hybridization present in this configuration is visualized in the envelope wavefunctions shown in Fig. 4l. From the same panel, we also note that, in the narrow well, the highest amplitude of the envelope wavefunctions is associated to $E_1$. Conversely, if we design the system with a $w_L$ value higher than the anticrossing condition as in 2267, where $w_L$= 16.0 nm and $b_t$ is still fixed at 2.3 nm, we observe, instead, that the maximum amplitude inside the narrow well corresponds to the envelope wavefunction associated to $E_2$ (Fig. 4n). This can be explained considering that, for $w_L$ being lower (higher) than the anticrossing value ($w_L \simeq$ 14 nm), $E_1$ ($E_2$) represents the ground state of the narrow well in the limit of uncoupled wells [33]. As a consequence, when $w_L$ is increased, the $E_2$ amplitude in the wide well lowers significantly. Since the wide well is doped and, as shown in the previous section, doping has a strong impact on the linewidths of the ISB absorption resonances, the change in the amplitude of the $E_2$ envelope wavefunction with $w_L$ may explain the strong reduction of the $\Gamma_2$ absorption linewidth when $w_L$ gets larger, observed in Fig. 4i. In addition, we note that, for $w_L \simeq$ 11.5 nm, the amplitude at the tunnel barrier interface of the $E_2$ envelope wavefunction is lower than that corresponding to the $E_1$ state, as shown in Fig. 4l. This is due to the fact that $E_2$ is the antibonding state forming as a result of the interwell interaction. Thus, the impact of the interface roughness scattering on the broadening of $\Gamma_2$ is expected to be lower with respect to $\Gamma_1$, well matching the behavior observed in Fig. 4i.

## 5. Conclusions

By combining THz spectroscopy with numerical calculations of the optical ISB absorption, we performed a detailed investigation of the dependence of quantum-confined electron states and ISB absorption resonances on the design of *n*-type Ge/SiGe ACQWs. The spectroscopic investigation was paralleled with an in-depth structural characterization obtained by merging STEM, XRD, and SIMS which showed excellent material quality and growth reproducibility.

Through a synergistic use of theory and experiment, we demonstrated an effective tuneability by design of the energy of ISB transitions as well as of the spatial overlap of the excited electronic wavefunctions. By varying the thickness of the tunneling barrier or the relative widths of the two coupled Ge wells, we extract a rich ensemble of information from optical experiments, such as the absorption-peak linewidths and their dependence of the ACQW design and doping concentration. The very good agreement between experiment and numerical calculations of the optical ISB absorptions highlights the accurate estimation of the *L*-point conduction-band offsets in the SiGe material system, for which we find a discontinuity of 117 meV for $x_{Ge}$= 0.81 and of 80 meV for $x_{Ge}$= 0.87, a range potentially suitable for operation of a silicon-based THz QCL.

**Funding:** This work is supported by the European Union research and innovation programme Horizon 2020 under grant no. 766719—FLASH project.

**References**

1. Bastard, G. *Wave Mechanics Applied to Semiconductor Heterostructures*; Wiley: New York, 1992.
2. Fox, M.; Ispasoiu, R. Quantum Wells, Superlattices, and Band-Gap Engineering. In *Springer Handbook of Electronic and Photonic Materials*, Kasap, S., Capper, P., Eds. Springer International Publishing: Cham, 2017; 10.1007/978-3-319-48933-9_40pp. 1-1.
3. Barnham, K.W.J.; Ballard, I.; Connolly, J.P.; Ekins-Daukes, N.J.; Kluftinger, B.G.; Nelson, J.; Rohr, C. Quantum well solar cells. *Physica E: Low-dimensional Systems and Nanostructures* **2002**, *14*, 27-36, doi:https://doi.org/10.1016/S1386-9477(02)00356-9.
4. Almosni, S.; Delamarre, A.; Jehl, Z.; Suchet, D.; Cojocaru, L.; Giteau, M.; Behaghel, B.; Julian, A.; Ibrahim, C.; Tatry, L., et al. Material challenges for solar cells in the twenty-first century: directions in emerging technologies. *Science and Technology of Advanced Materials* **2018**, *19*, 336-369, doi:10.1080/14686996.2018.1433439.
5. Welser, R.E.; Polly, S.J.; Kacharia, M.; Fedorenko, A.; Sood, A.K.; Hubbard, S.M. Design and Demonstration of High-Efficiency Quantum Well Solar Cells Employing Thin Strained Superlattices. *Scientific Reports* **2019**, *9*, 13955, doi:10.1038/s41598-019-50321-x.
6. Köhler, R.; Tredicucci, A.; Beltram, F.; Beere, H.E.; Linfield, E.H.; Davies, A.G.; Ritchie, D.A.; Iotti, R.C.; Rossi, F. Terahertz semiconductor-heterostructure laser. *Nature* **2002**, *417*, 156-159, doi:10.1038/417156a.
7. Williams, B.S. Terahertz quantum-cascade lasers. *Nat. Photonics* **2007**, *1*, 517-525, doi:10.1038/nphoton.2007.166.
8. Vitiello, M.S.; Scalari, G.; Williams, B.; De Natale, P. Quantum cascade lasers: 20 years of challenges. *Opt. Express* **2015**, *23*, 5167-5182, doi:10.1364/OE.23.005167.
9. Giorgioni, A.; Paleari, S.; Cecchi, S.; Vitiello, E.; Grilli, E.; Isella, G.; Jantsch, W.; Fanciulli, M.; Pezzoli, F. Strong confinement-induced engineering of the g factor and lifetime of conduction electron spins in Ge quantum wells. *Nature Communications* **2016**, *7*, 13886, doi:10.1038/ncomms13886.
10. Hendrickx, N.W.; Franke, D.P.; Sammak, A.; Scappucci, G.; Veldhorst, M. Fast two-qubit logic with holes in germanium. *Nature* **2020**, *577*, 487-491, doi:10.1038/s41586-019-1919-3.
11. Faist, J.; Capasso, F.; Sivco, D.L.; Sirtori, C.; Hutchinson, A.L.; Cho, A.Y. Quantum Cascade Laser. *Science* **1994**, *264*, 553, doi:10.1126/science.264.5158.553.
12. Gauthier-Lafaye, O.; Julien, F.H.; Cabaret, S.; Lourtioz, J.M.; Strasser, G.; Gornik, E.; Helm, M.; Bois, P. High-power GaAs/AlGaAs quantum fountain unipolar laser emitting at 14.5 μm with 2.5% tunability. *Appl. Phys. Lett.* **1999**, *74*, 1537-1539, doi:10.1063/1.123608.
13. Paul, D.J. The progress towards terahertz quantum cascade lasers on silicon substrates. *Laser & Photonics Reviews* **2010**, *4*, 610-632, doi:10.1002/lpor.200910038.
14. Grange, T.; Stark, D.; Scalari, G.; Faist, J.; Persichetti, L.; Di Gaspare, L.; De Seta, M.; Ortolani, M.; Paul, D.J.; Capellini, G., et al. Room temperature operation of n-type Ge/SiGe terahertz quantum cascade lasers predicted by non-equilibrium Green's functions. *Appl. Phys. Lett.* **2019**, *114*, 111102, doi:10.1063/1.5082172.
15. Sabbagh, D.; Schmidt, J.; Winnerl, S.; Helm, M.; Di Gaspare, L.; De Seta, M.; Virgilio, M.; Ortolani, M. Electron Dynamics in Silicon–Germanium Terahertz Quantum Fountain Structures. *ACS Photonics* **2016**, *3*, 403-414, doi:10.1021/acsphotonics.5b00561.

16. Ciano, C.; Virgilio, M.; Bagolini, L.; Baldassarre, L.; Pashkin, A.; Helm, M.; Montanari, M.; Persichetti, L.; Di Gaspare, L.; Capellini, G., et al. Terahertz Absorption-Saturation and Emission from Electron-doped Germanium Quantum Wells. *Opt. Express* **2020**, *(in press)*, doi:https://doi.org/10.1364/OE.381471.
17. Sun, G.; Cheng, H.H.; Menéndez, J.; Khurgin, J.B.; Soref, R.A. Strain-free Ge/GeSiSn quantum cascade lasers based on L-valley intersubband transitions. *Appl. Phys. Lett.* **2007**, *90*, 251105, doi:10.1063/1.2749844.
18. Sun, G.; Soref, R.A.; Cheng, H.H. Design of a Si-based lattice-matched room-temperature GeSn/GeSiSn multi-quantum-well mid-infrared laser diode. *Opt. Express* **2010**, *18*, 19957-19965, doi:10.1364/OE.18.019957.
19. Virgilio, M.; Ortolani, M.; Teich, M.; Winnerl, S.; Helm, M.; Sabbagh, D.; Capellini, G.; De Seta, M. Combined effect of electron and lattice temperatures on the long intersubband relaxation times of Ge/Si$_x$Ge$_{1-x}$ quantum wells. *Phys. Rev. B* **2014**, *89*, 045311, doi:10.1103/PhysRevB.89.045311.
20. Ciano, C.; Virgilio, M.; Bagolini, L.; Baldassarre, L.; Rossetti, A.; Pashkin, A.; Helm, M.; Montanari, M.; Persichetti, L.; Di Gaspare, L., et al. Electron Population Dynamics in Optically Pumped Asymmetric Coupled Ge/SiGe Quantum Wells: Experiment and Models. *Photonics* **2019**, *7*, 2.
21. Ciano, C.; Virgilio, M.; Montanari, M.; Persichetti, L.; Di Gaspare, L.; Ortolani, M.; Baldassarre, L.; Zoellner, M.H.; Skibitzki, O.; Scalari, G., et al. Control of Electron-State Coupling in Asymmetric Ge/SiGe Quantum Wells. *Physical Review Applied* **2019**, *11*, 014003, doi:10.1103/PhysRevApplied.11.014003.
22. Driscoll, K.; Paiella, R. Design of n-type silicon-based quantum cascade lasers for terahertz light emission. *J. Appl. Phys.* **2007**, *102*, 093103, doi:10.1063/1.2803896.
23. Valavanis, A.; Dinh, T.V.; Lever, L.J.M.; Ikonić, Z.; Kelsall, R.W. Material configurations for $n$-type silicon-based terahertz quantum cascade lasers. *Phys. Rev. B* **2011**, *83*, 195321, doi:10.1103/PhysRevB.83.195321.
24. Montanari, M.; Virgilio, M.; Manganelli, C.L.; Zaumseil, P.; Zoellner, M.H.; Hou, Y.; Schubert, M.A.; Persichetti, L.; Di Gaspare, L.; De Seta, M., et al. Photoluminescence study of interband transitions in few-layer, pseudomorphic, and strain-unbalanced Ge/GeSi multiple quantum wells. *Phys. Rev. B* **2018**, *98*, 195310, doi:10.1103/PhysRevB.98.195310.
25. De Seta, M.; Capellini, G.; Ortolani, M.; Virgilio, M.; Grosso, G.; Nicotra, G.; Zaumseil, P. Narrow intersubband transitions in n-type Ge/SiGe multi-quantum wells: control of the terahertz absorption energy trough the temperature dependent depolarization shift. *Nanotechnol.* **2012**, *23*, 465708, doi:10.1088/0957-4484/23/46/465708.
26. Helm, M. The basic physics of intersubband transitions. In *Intersubband Transition in Quantum Wells: Physics and Device Applications I. Semiconductors and Semimetals*, Liu, H.C., Ed. Academic: San Diego, 2000; pp. 1-99.
27. Busby, Y.; De Seta, M.; Capellini, G.; Evangelisti, F.; Ortolani, M.; Virgilio, M.; Grosso, G.; Pizzi, G.; Calvani, P.; Lupi, S., et al. Near- and far-infrared absorption and electronic structure of Ge-SiGe multiple quantum wells. *Phys. Rev. B* **2010**, *82*, 205317, doi:doi: 10.1103/PhysRevB.82.205317.
28. Tersoff, J. Dislocations and strain relief in compositionally graded layers. *Appl. Phys. Lett.* **1993**, *62*, 693-695, doi:10.1063/1.108842.
29. Montalenti, F.; Rovaris, F.; Bergamaschini, R.; Miglio, L.; Salvalaglio, M.; Isella, G.; Isa, F.; Von Känel, H. Dislocation-Free SiGe/Si Heterostructures. *Crystals* **2018**, *8*, 257.

30. Grange, T.; Mukherjee, S.; Capellini, G.; Montanari, M.; Persichetti, L.; Di Gaspare, L.; Birner, S.; Attiaoui, A.; Moutanabbir, O.; Virgilio, M., et al. Atomic-scale insights into diffuse heterointerfaces: from three-dimensional roughness analysis to a generalized theory of interface scattering. *submitted* **2020**.

31. Capellini, G.; De Seta, M.; Zaumseil, P.; Kozlowski, G.; Schroeder, T. High temperature x ray diffraction measurements on Ge/Si(001) heterostructures: A study on the residual tensile strain. *J. Appl. Phys.* **2012**, *111*, 073518, doi:doi: 10.1063/1.3702443.

32. Virgilio, M.; Sabbagh, D.; Ortolani, M.; Di Gaspare, L.; Capellini, G.; De Seta, M. Physical mechanisms of intersubband-absorption linewidth broadening in *s*-Ge/SiGe quantum wells. *Phys. Rev. B* **2014**, *90*, 155420, doi:10.1103/PhysRevB.90.155420.

33. Among our samples, the lowest interwell interaction is obtained in 2224, which features the thickest tunnel barrier, and where, consistently with wL= 12.0 nm, we observe the strongest amplitude of E1 in the narrow well, see Fig. 4m.